\documentclass[lettersize,journal]{IEEEtran}
\usepackage{amsmath,amsfonts}
\usepackage{algorithmic}
\usepackage{algorithm}
\usepackage{array}
\usepackage[caption=false,font=footnotesize,labelfont=rm,textfont=rm]{subfig}
\usepackage{textcomp}
\usepackage{stfloats}
\usepackage{url}
\usepackage{verbatim}
\usepackage{graphicx}
\usepackage{cite}
\usepackage{cleveref}
\usepackage{url}
\usepackage{textcomp}

\begin{document}

\title{A Digital and Compact High-Precision Locking System for Pulse Laser Repetition Frequency}

\author{Qibin~Zheng,
Zhengyi~Tao,
Lei~Wang,
Zhaohui~Bu,
Liguo~Zhou,
Zuanming~Jin,
Zhao~Wang%
\thanks{ Manuscript received xxx xx, 2024; revised xxx xx, 2024. This project is supported in part by grants from National Natural Science Foundation of China (No. 12105177 and 11904423) 
and a grant from National Key R\&D Program of China (No. 2023YFF0719200). (Corresponding author: Qibin Zheng and Zhao Wang).}
\thanks{ Qibin Zheng, Zhengyi Tao, Lei Wang, Zhaohui Bu, Liguo Zhou are with the Quantum Medical Sensing Laboratory and School of Health Science and Engineering, University of Shanghai for Science and Technology, Shanghai 200093, China (e-mail: qbzheng@usst.edu.cn).

Zuanming Jin is with the Terahertz Technology Innovation Research Institute, University of Shanghai for Science and Technology, Shanghai 200093, China.

Zhao Wang is with the Southern University of Science and Technology, Shenzhen 518055, China, and  Shenzhen International Quantum Academy, Shenzhen 518048, China (e-mail: wangz7@sustech.edu.cn).
}}


\markboth{IEEE TRANSACTIONS ON INSTRUMENTATION AND MEASUREMENT Class Files,~Vol.~XX, No.~X, November~2024}%
{Shell \MakeLowercase{\textit{et al.}}: A Sample Article Using IEEEtran.cls for IEEE Journals}


\IEEEpubid{0000--0000/00\$00.00~\copyright~2021 IEEE}

\maketitle

\begin{abstract}
  This paper proposes a novel approach that employs error amplification and ADC-based dual-mixer time-difference (ADC-based-DMTD) technique 
  for high-precision locking of laser repetition frequency with compact size. 
  This electronic system consists of two main components: a digitized error amplification module (EAM) 
  and an FPGA-based digital frequency locking module (DFLM). 
  The EAM mainly integrates a configurable frequency generator (CFG), a configurable frequency multiplier (CFM) and a mixer 
  to process the laser pulses and a high-stability reference source (e.g., an atomic clock), 
  enabling high-precision locking of pulse lasers operating at different repetition frequencies.
  By employing frequency multiplication and mixing, the EAM amplifies the laser's frequency error and performs frequency down-conversion, 
  enhancing measurement sensitivity and reducing the hardware requirements of the back-end. 
  The DFLM receives the EAM outputs by using an ADC-based-DMTD method to precisely measure frequency errors, 
  then the digital proportional-integral-derivative (PID) controller provides feedback to achieve accurate frequency locking.
  Initial testing with a voltage-controlled oscillator (VCO) demonstrated excellent locking performance, 
  achieving an Allan deviation of $9.58 \times 10^{-14}$ at 10 seconds and a standard deviation (STD) of 7.7 \textmu Hz root mean square (RMS) after locking, 
  marking a five-order-of-magnitude stability enhancement. 
  In laboratory experiments with a custom-built femtosecond fiber laser, 
  the system achieved robust locking of the repetition frequency, 
  with a stability improvement from $1.51 \times 10^{-7}$ to $1.12 \times 10^{-12}$ at a 10-second gate time and an STD of 0.43 mHz RMS after locking.
\end{abstract}


\section{Introduction} \label{Introduction}
\IEEEPARstart{P}{ulse} lasers characterized by narrow-lines and high repetition frequency stability are essential components in optical frequency combs. 
They have widespread applications in both industrial settings and scientific research,
including high-precision measurement\cite{WOS:000532836000026}, \cite{10.1117/1.OE.54.8.084112}, optical communication\cite{WOS:001000552100456}, \cite{Kim:16}, 
high-resolution optical spectroscopy\cite{electronics12132762}, \cite{LIU2022106900}, etc.
Optical frequency combs are advanced tools that generate millions of mutually coherent and evenly spaced modes, 
enabling precise phase-coherent connections across the optical, microwave, and terahertz frequency ranges. 
These frequency combs are essential for a variety of applications and fundamental scientific research\cite{WOS:000539031100001}, \cite{10.1063/1.4928163}.
However, optic combs have been limited by various factors such as strain, mechanical perturbations\cite{2011Mechanisms}, temperature and humidity fluctuations\cite{2015Repetition}. 
These environmentally induced variations can cause repetition frequency drift in the comb, leading to instability and inaccuracy in measurements\cite{doi:10.1126/science.aay3676}, \cite{4346524}. 
The frequency instability, particularly the resulting jitter in the laser, can significantly degrade measurement resolution.
Furthermore, many experiments often need to be performed outdoors, where environmental control is limited, and system size is strictly constrained\cite{WOS:000481889000034}, \cite{WOS:000382012100007}, \cite{Dinkelaker:17}.
Therefore, a highly integrated, robust and easy-tunable system which can stabilize the laser repetition frequency with high precession is very advantageous and necessary in these applications.
\IEEEpubidadjcol

Methods such as analog phase-locked loop (PLL) and charge pump PLL (CP-PLL) 
have been employed for laser repetition frequency stabilization\cite{6172209}, \cite{Zhang2021StudyOU} in past few decades.
However, these methods are constrained by environmental sensitivity, feedback accuracy, and limited tunability.
Since laser experiments are often dynamic and complex, 
achieving precise and flexible frequency locking with configurable parameters 
necessitates the use of advanced digital systems.
Digital systems based on microcontroller units (MCUs) and FPGAs
have emerged as preferred approaches for laser frequency stabilization\cite{10.1063/1.4903869}, \cite{10.1063/1.3455830}. 
While MCUs typically offer bandwidths in the kHz range\cite{10.1063/5.0090384}, 
FPGAs provide higher bandwidth and are better suited for applications requiring greater precision and flexibility\cite{10.1063/5.0050999}. 

The typical FPGA-based platforms like the Red Pitaya STEMlab have been increasingly adopted for real-world applications\cite{8087380}, \cite{WOS:001010633300001}, \cite{10.1063/1.5083797}.
Some systems use fully digital FPGA architectures for frequency measurement and feedback \cite{WOS:001029190800017}, \cite{WOS:000497661500024}, \cite{MAJUMDER2024110247}, 
which eliminates interference from analog components while offering excellent timing resolution and high bandwidth.
But lasers typically operate at repetition frequencies in the tens to hundreds of MHz, or even GHz \cite{Paschottapulse_repetition_rate}, 
making direct signal capture with ultra-high-speed ADCs prohibitively expensive. 
To address this issue, common approaches include frequency divider or higher-order harmonics mixing to lower the input signal frequency\cite{10.1063/1.4928163}, \cite{2015Repetition}. 
Frequency dividers reduce the repetition frequency error, leading to a loss in measurement accuracy. 
The higher-order harmonics mixing method offering a solution for improving measurement accuracy 
for it contains greater frequency error as the higher harmonics of the laser increase.
However, the accompanying reduction in energy results in a lower signal-to-noise ratio, 
making them highly susceptible to environmental influences.

\begin{figure*}[t]
  \centering
  \includegraphics[width=0.85\linewidth]{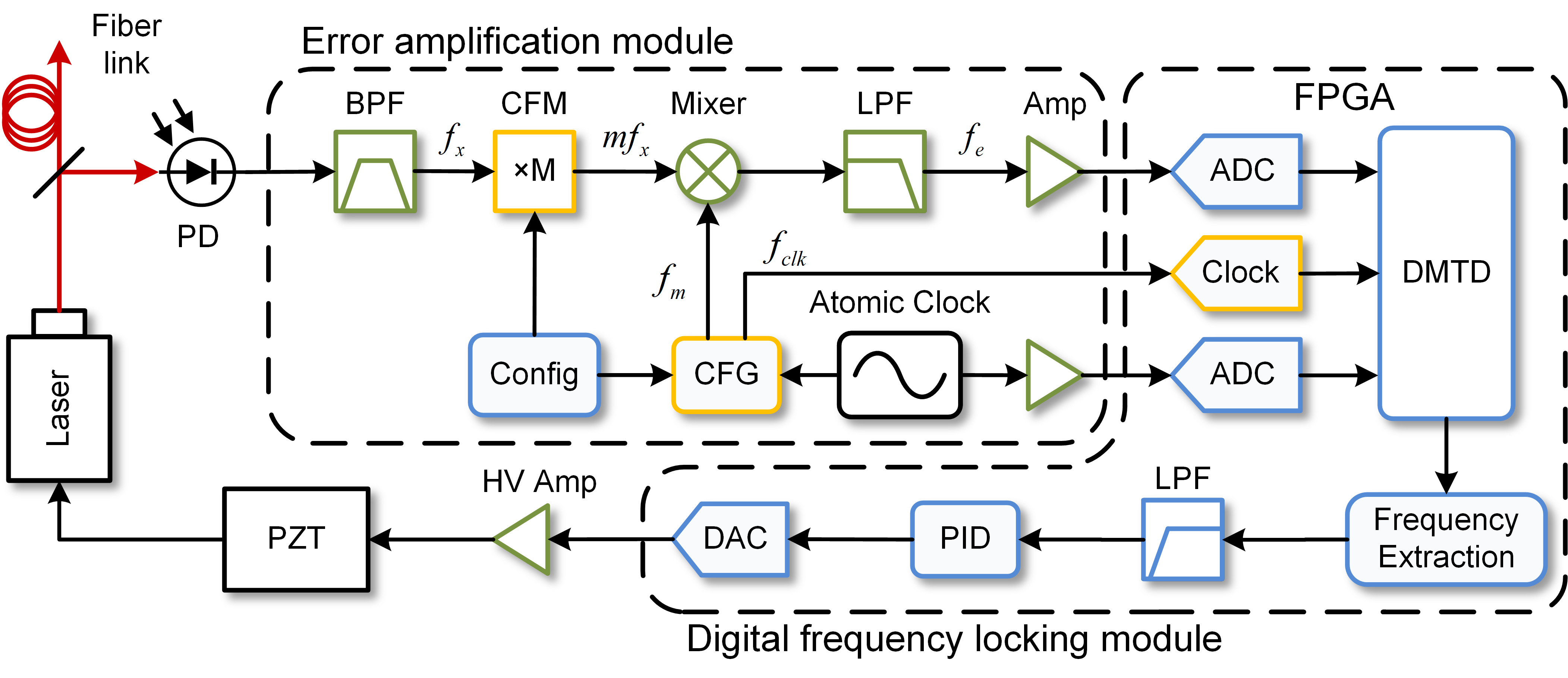}
  \caption{Schematic of locking circuits for laser frequency stabilization. PD: photodiode, BPF: band-pass filter, LPF: low-pass filter, 
  CFM: configurable frequency multiplier, CFG: configurable frequency generator, PZT: piezoelectric transducer, HV: high voltage, 
  Amp: voltage amplifier, ADC: analog-to-digital converter, DMTD: dual-mixer time-difference, DAC: digital-to-analog converter, PID: proportional integral derivative.}
  \label{fig:overall}
  \end{figure*}
In this paper, we present an FPGA-based, high-precision system for laser repetition frequency locking. 
The system employs digital error amplification and ADC-based dual-mixer time-difference (ADC-based-DMTD) technique to
achieve accurate measurement.
The digital error amplification module (EAM) uses digital PLL chip to multiply the repetition frequency, 
which multiplies the frequency errors while down-converting the signal to simplify the requirements of back-end hardware. 
The approach enhances sensitivity and provides flexibility for adapting to different repetition frequencies.
The digital frequency locking module (DFLM) captures the high-stability reference source and error signals via high-speed, high-precision ADCs. 
By referencing the DMTD method \cite{1986Measurements}, \cite{6831788}, 
DFLM utilizes an ADC-based-DMTD method to perform high-precision phase measurements.
Followed by frequency extraction and filtering, this processed signal is sent to a PID controller, which adjusts the DAC to achieve real-time feedback, 
ensuring precise laser frequency locking and long-term stability.
Both the EAM and DFLM are closely linked with the FPGA, 
allowing for easy system parameter adjustments to adapt to different laser repetition frequencies.
By utilizing these methods, the system improves measurement accuracy while maintaining a highly integrated structure,
and can provide robust laser stabilization across a wide range of repetition frequencies, 
suitable for diverse applications.

The paper is organized as follows: 
Section \ref{Methodology} gives an overview of the laser repetition frequency locking system, 
detailing the system design and theory analysis.
Section \ref{Implementation and Performance} describes the implementation of the electronic prototype 
and presents performance evaluations through noise floor testing and VCO stability experiments.
A laser stability experiment is conducted to analyze long-term frequency fluctuations 
and Allan deviation before and after locking.
A summary of this research and the discussion of future improvements are given in Section \ref{Conclusion}.

\section{Methodology} \label{Methodology}
\subsection{Overall Design}
The schematic of the locking circuits for laser frequency stabilization is illustrated in Fig.\ref{fig:overall}. 
This system measures the frequency difference between the laser repetition frequency and a high-stability reference source, 
an atomic clock in this case, enabling precise locking of the laser frequency. 
The pulse of the laser is first converted to electrical signal using a photodiode (PD). 
This signal is then processed by the EAM, which amplifies and down-converts the frequency error.
The configurable frequency multiplier (CFM) performs frequency multiplication on the laser signal, 
producing high-frequency signals for mixing. 
The configurable frequency generator (CFG) uses the atomic clock to produce a stable high frequency signal also for mixing.
After mixing, a low-pass filter (LPF) is used to obtain the amplified error signal.
Through this process, the frequency error is amplified while the frequency of the signal to be sampled is lowered, 
thus reducing the requirements of the back-end circuitry. 

The error signal and atomic clock signal are sampled by high-speed ADCs for FPGA processing.
An ADC-based-DMTD algorithm is adopted to calculate the phase difference between the sampled signals.
Then this phase difference is converted into a frequency difference, passed through a low-pass filter, 
and fed to the PID module, which controls a high-speed, high-precision DAC for feedback.
Depending on the laser's repetition frequency and the feedback devices employed,
CFM, CFG and digital units within the FPGA can be configured to accommodate them.
This capability facilitates the configuration of the EAM and the digital units within the FPGA, 
ensuring the system's adaptability and robustness.

\subsection{Design of the digitized Error Amplification Module (EAM)}
Advancements in technology have significantly improved the sampling rates and resolution of ADCs, 
making fully digital laser repetition frequency locking systems based on high-speed ADCs feasible\cite{10.1063/1.5083797}.
In applications such as optical frequency combs, laser repetition frequencies can reach tens to hundreds of MHz, or even GHz\cite{Paschottapulse_repetition_rate}. 
While ultra-high-speed ADCs can directly capture these signals, 
this approach is costly and introduces considerable design complexity.
To accommodate lasers with higher repetition frequencies, 
common techniques involve down-converting using methods such as frequency divider and high-order harmonic mixing\cite{10.1063/1.4928163}, \cite{2015Repetition}. 
Frequency divider lowers the frequency of the error signal during down-conversion, which can lead to precision loss.
High-order harmonic mixing improves accuracy by extracting higher harmonics of the laser signal, 
which contains higher frequency error signal.
But the narrow pass band BFP used to extract higher harmonics often suffer from poor out-of-band rejection, 
increased costs, and potential spectral aliasing, all of which can compromise measurement accuracy. 
Additionally, as harmonic order increases, signal amplitude decreases, necessitating high-gain amplification. 
This amplification, while boosting the signal, introduces noise, making high-precision measurements more challenging.

To improve precision while reducing hardware costs, 
a digitized EAM is designed to preprocesses the laser signal by incorporating principles of frequency multiplication and beat frequency methods. 
The PLL-based CFM multiplies the laser repetition frequency, while a PLL-based CFG, equipped with jitter-cleaning functionality, 
generates stable reference signals from the atomic clock for mixing and provides clock signals to the DFLM. 
The digital method for frequency multiplies ensures the energy of high-frequency signals while providing strong resistance to interference.
By amplifying and down-converting the laser repetition frequency error using the digitized EAM, 
measurement precision is enhanced while hardware costs are reduced.

The input laser repetition frequency signal undergoes band-pass filtering to isolate the desired repetition signal:
\begin{equation}
  \label{equationEAM1}
  {{f}_{x}}={{f}_{0}}+\Delta f
\end{equation}
where ${{f}_{0}}$ is the expected laser repetition frequency, 
and $\Delta f$ is the frequency difference between the expected and actual laser repetition frequency.

The CFM multiplies the laser repetition frequency by a factor of $m$, generating a high-frequency signal $m{{f}_{x}}$.

Simultaneously, the CFG processes the stable reference signal ${{f}_{r}}$ from the atomic clock to produce a stable mixing signal:
\begin{equation}
  \label{equationEAM2}
  {{f}_{m}}=m{{f}_{0}}-{{f}_{r}}.
\end{equation}

The high frequency input signal $m{{f}_{x}}$ and the stable mixing signal ${{f}_{m}}$ are then mixed and passed through an LPF, producing an amplified error signal:
\begin{equation}
  \label{equationEAM3}
  {{f}_{e}}={{f}_{r}}+m\Delta f.
\end{equation}

This amplified error signal, which matches the reference frequency, reduces the impact of back-end noise, thereby improving measurement precision.

The EAM can be configured to accommodate different laser repetition frequencies 
by adjusting the multiplication factor and the frequency of the mixing signal generated by the PLL, 
ensuring the processed signal can be effectively sampled by the ADC.

\begin{figure}[t]
  \centering
  \subfloat[]{
    \includegraphics[scale=0.6]{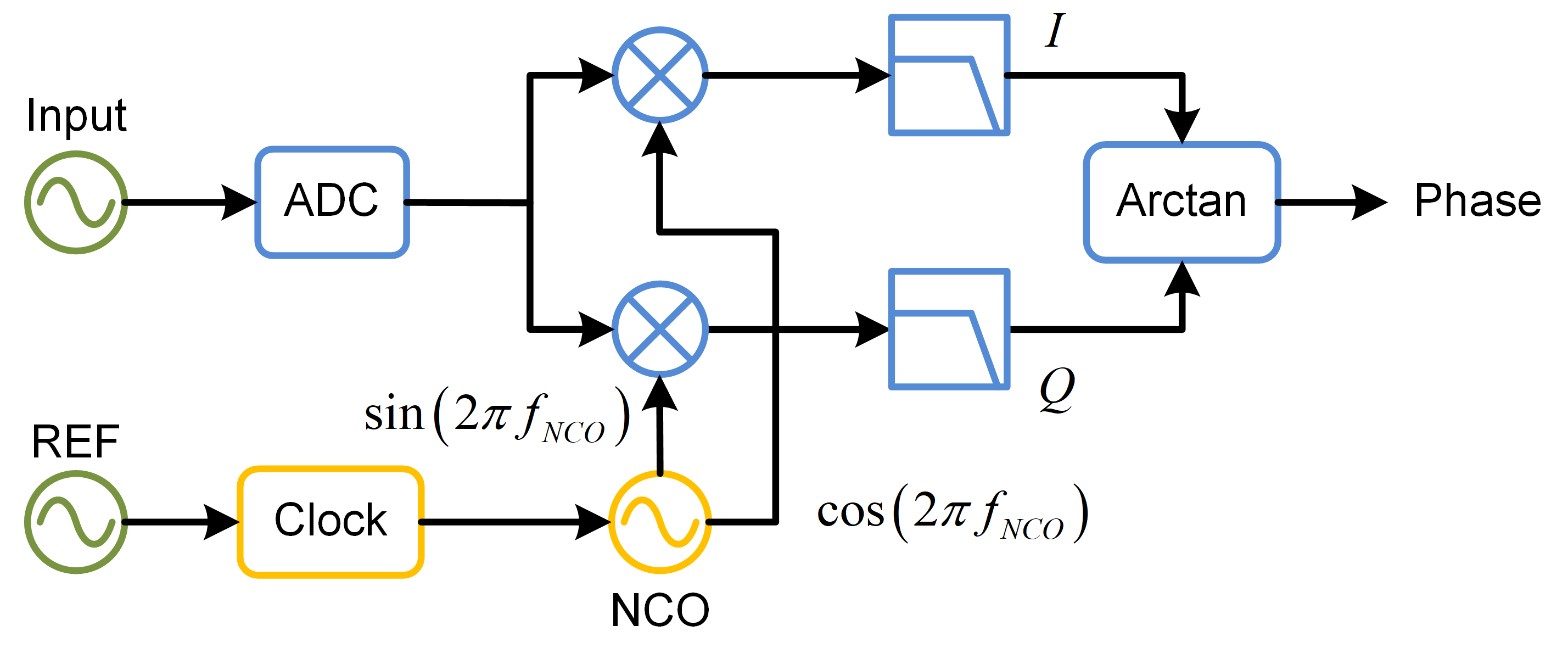}}\\
    \label{fig:DQD}
    \subfloat[]{
    \label{fig:DMTD}
    \includegraphics[scale=0.6]{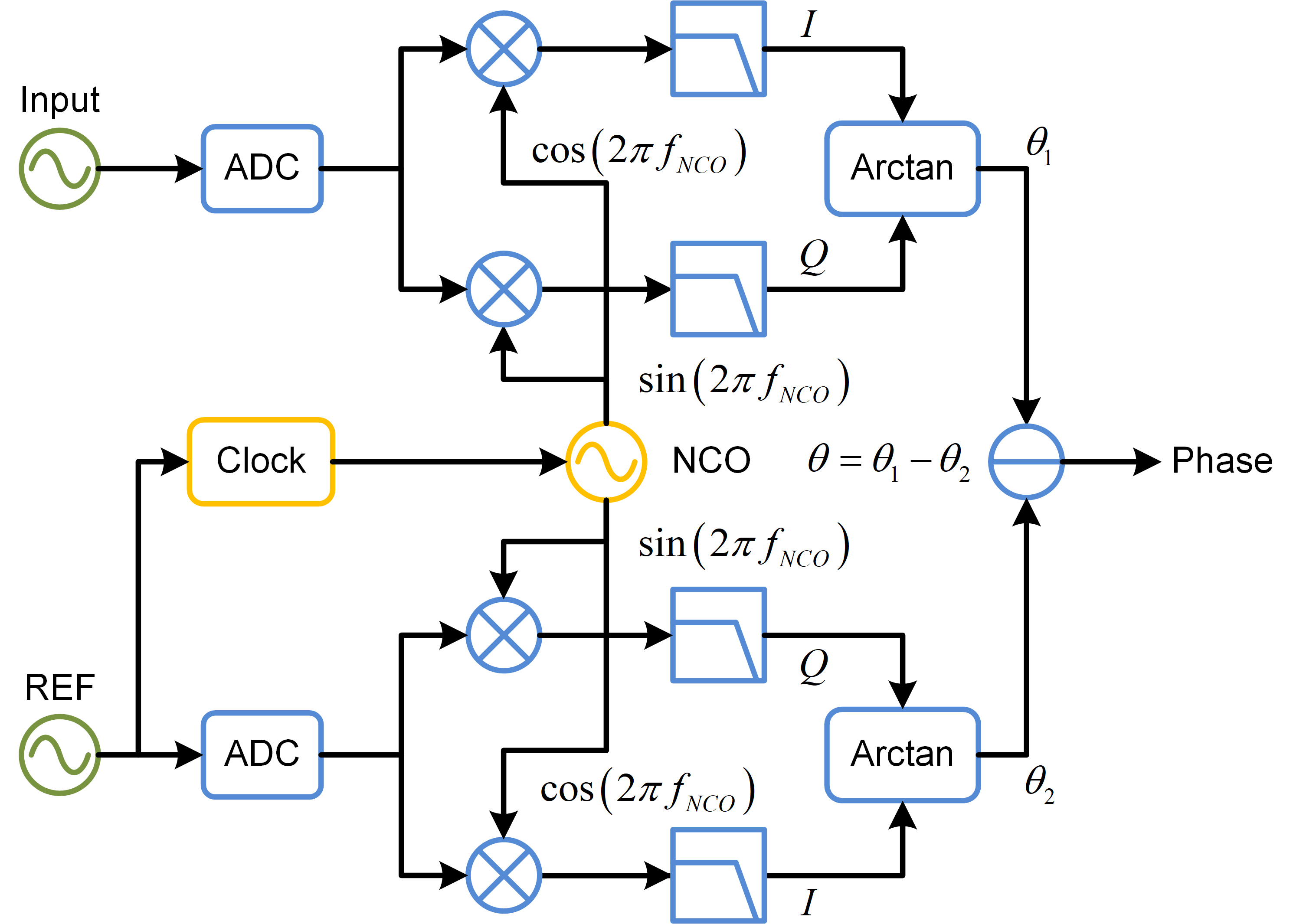}}
  \caption{Block diagram of measurement methods. (a) digital quadrature demodulation (DQD) method (b) ADC-based-DMTD method. 
  REF: reference source, NCO: numerically controlled oscillator.}
  \label{fig:DMTD_DQD}
\end{figure}
\subsection{Design of ADC-based-DMTD and Frequency Extraction Module}
In traditional laser repetition frequency locking systems, the digital quadrature demodulation (DQD) method \cite{Tourigny2018An}, \cite{10.1063/1.4928163}, 
shown in Fig.\ref{fig:DMTD_DQD}(a), is widely used for frequency measurement due to its simple structure and easy-tunable ability. 
However, it is susceptible to inaccuracies caused by the instability and phase noise from the numerically controlled oscillator (NCO). 
In contrast, DMTD method\cite{1986Measurements}, \cite{6831788} can counteract noise from the NCO, 
offering higher accuracy while maintaining flexible sampling period\cite{1983Automated}.

The traditional analog DMTD method measures the phase difference of beat signals 
through zero-crossing detection and time interval counters \cite{1418555}, \cite{2000Common}. 
However, the reliance on analog components introduces additional noise into the system \cite{1418555}. 
While the digital DMTD (D-DMTD) \cite{5556289} approach eliminates analog noise by using fully digital techniques, 
both analog and digital methods are susceptible to noise interference at zero-crossing points, leading to measurement errors. 
Additionally, these methods adjust the measurement period by altering the beat frequency \cite{S2009Novel}, 
which can lower resolution when the measurement period is shortened.

This paper employs an ADC-based-DMTD method Fig.\ref{fig:DMTD_DQD}(b) to measure the frequency difference between two signals. 
This method avoids the zero-crossing noise issue associated with analog methods and utilizes digital components, 
such as mixers and filters, which offer greater consistency\cite{10.1063/1.4950898}, \cite{4303427}. 
In the symmetric design of the ADC-based-DMTD  method, this consistency maximizes the cancellation of common-mode noise, 
significantly reducing noise and enhancing measurement accuracy.
Additionally, the method does not require the two signals to have identical nominal frequencies, 
enabling it to measure signals with substantial frequency deviations in the MHz range. 
This feature broadens the system's measurement tolerance.

The process begins by sampling both the amplified error signal and the reference signal using an ADC. 
The amplified error frequency signal is represented as:
\begin{equation}
\label{equation1}
{{U}_{e}}\left( t \right)={{A}_{e}}sin\left[ 2\pi {{f}_{e}}t+{{\varphi }_{e}}+\Delta {{\varphi }_{e}}(t) \right]
\end{equation}
while the reference signal from a highly stable source is expressed as:
\begin{equation}
  \label{equation2}
  {{U}_{r}}\left( t \right)={{A}_{r}}sin[2\pi {{f}_{r}}t+{{\varphi }_{r}}+\Delta {{\varphi }_{r}}(t)].
\end{equation}

Here, ${{A}_{e}}$ and ${{A}_{r}}$ denote the amplitudes, ${{f}_{e}}$ and ${{f}_{r}}$ represent the frequencies, 
and ${{\varphi }_{e}}$, ${{\varphi }_{r}}$ are the initial phases. 
The terms $\Delta {{\varphi }_{e}}(t)$ and $\Delta {{\varphi }_{r}}(t)$ indicate the random phase noise in the signals.

Two orthogonal signals are then generated using the NCO within the FPGA:
\begin{equation}
  \label{equation3}
  {{U}_{s1}}\left( t \right)={{A}_{s1}}sin\left( 2\pi {{f}_{s1}}t+{{\varphi }_{s1}}+\Delta {{\varphi }_{s1}}(t) \right)
\end{equation}
\begin{equation}
  \label{equation4}
  {{U}_{s2}}\left( t \right)={{A}_{s2}}cos\left( 2\pi {{f}_{s2}}t+{{\varphi }_{s2}}+\Delta {{\varphi }_{s2}}(t) \right).
\end{equation}

Since the two orthogonal signals are generated by the same NCO, ideally, 
they satisfy ${{A}_{s1}}={{A}_{s2}}$, ${{f}_{s1}}={{f}_{s2}}={{f}_{NCO}}$, and ${{\varphi }_{s1}}={{\varphi }_{s2}}$. 
The random phase jitter contained in the orthogonal signals, 
represented by $\Delta {{\varphi }_{s1}}\left( t \right)$ and $\Delta {{\varphi }_{s2}}\left( t \right)$, 
is also symmetric and satisfies $\Delta {{\varphi }_{s1}}\left( t \right)=\Delta {{\varphi }_{s2}}\left( t \right)$ under ideal conditions.

Next, the reference signal ${{U}_{r}}$ is digitally mixed with the NCO signals and low-pass filtered, 
yielding digital signals with low-frequency components:
\begin{equation}
  \label{equation5}
  \resizebox{0.9\hsize}{!}{${{I}_{1}}\left( t \right)\approx {{A}_{I1}}\sin \left[ 2\pi {{f}_{h1}}t-{{\varphi }_{s2}}+{{\varphi }_{r}}-\Delta {{\varphi }_{s2}}(t)+\Delta {{\varphi }_{r}}(t) \right]$}
\end{equation}
\begin{equation}
  \label{equation6}
  \resizebox{0.9\hsize}{!}{${{Q}_{1}}\left( t \right)\approx {{A}_{Q1}}\cos \left[ 2\pi {{f}_{h1}}t-{{\varphi }_{s1}}+{{\varphi }_{r}}-\Delta {{\varphi }_{s1}}(t)+\Delta {{\varphi }_{r}}(t) \right]$}.
\end{equation}

Here, ${{f}_{h1}}={{f}_{r}}-{{f}_{s1,2}}$. The same operations are applied to the amplified error frequency signal ${{U}_{e}}$:
\begin{equation}
  \label{equation7}
  \resizebox{0.9\hsize}{!}{${{I}_{2}}\left( t \right)\approx {{A}_{I2}}\sin \left[ 2\pi {{f}_{h2}}t-{{\varphi }_{s2}}+{{\varphi }_{e}}-\Delta {{\varphi }_{s2}}(t)+\Delta {{\varphi }_{e}}(t) \right]$}
\end{equation}
\begin{equation}
  \label{equation8}
  \resizebox{0.9\hsize}{!}{${{Q}_{2}}\left( t \right)\approx {{A}_{Q2}}\cos \left[ 2\pi {{f}_{h2}}t-{{\varphi }_{s1}}+{{\varphi }_{e}}-\Delta {{\varphi }_{s1}}(t)+\Delta {{\varphi }_{e}}(t) \right]$}
\end{equation}
where ${{f}_{h2}}={{f}_{e}}-{{f}_{s1,2}}$. 

The phase difference is derived from demodulating IQ signals as:
\begin{equation}
  \label{equation9}
  \begin{aligned}
  {{\theta }_{1}}\left( t \right)& ={{\tan }^{-1}}\frac{{{I}_{1}}\left( t \right)}{{{Q}_{1}}\left( t \right)} \\
  & =2\pi {{f}_{h1}}t-{{\varphi }_{s1}}+{{\varphi }_{r}}-\Delta {{\varphi }_{s1}}+\Delta {{\varphi }_{r}}
  \end{aligned}
\end{equation}
\begin{equation}
  \label{equation10}
  \begin{aligned}
  {{\theta }_{2}}\left( t \right)& ={{\tan }^{-1}}\frac{{{I}_{2}}\left( t \right)}{{{Q}_{2}}\left( t \right)} \\
  & =2\pi {{f}_{h2}}t-{{\varphi }_{s2}}+{{\varphi }_{e}}-\Delta {{\varphi }_{s1}}+\Delta {{\varphi }_{e}}
\end{aligned}
\end{equation}
where ${{\theta }_{1}}\left( t \right)$ is the phase difference between the reference signal ${{U}_{r}}$ and the two orthogonal signals generated by the NCO. 
Similarly, ${{\theta }_{2}}\left( t \right)$ represents the phase difference between the amplified error frequency signal ${{U}_{e}}$ and the two orthogonal signals generated by the NCO.

By subtracting ${{\theta }_{2}}(t)$ from ${{\theta }_{1}}(t)$, 
the common-mode noise $\Delta {{\varphi }_{s1}}\left( t \right)$ from the NCO was canceled.
The phase difference between the amplified error frequency signal and the reference signal is calculated as:
\begin{equation}
  \label{equation11}
  \begin{aligned}
  \theta \left( t \right)& ={{\theta }_{1}}(t)-{{\theta }_{2}}(t) \\
  & =2\pi \left( {{f}_{r}}-{{f}_{e}} \right)t+{{\varphi }_{r}}-{{\varphi }_{e}}+\Delta {{\varphi }_{r}}\left( t \right)-\Delta {{\varphi }_{e}}\left( t \right).
\end{aligned}
\end{equation}

The frequency difference between the reference signal and the amplified error signal can be calculated by finding the derivative of phase difference with time $t$:
\begin{equation}
  \label{equation12}
  \begin{aligned}
  {{f}_{r,e}}& =\frac{\theta {{\left( t \right)}^{\prime }}}{2\pi }={{f}_{r}}-{{f}_{e}}+\frac{\Delta {{\varphi }_{r}}{{\left( t \right)}^{\prime }}-\Delta {{\varphi }_{e}}{{\left( t \right)}^{\prime }}}{2\pi } \\
  & =m\Delta f+\frac{\Delta {{\varphi }_{r}}{{\left( t \right)}^{\prime }}-\Delta {{\varphi }_{e}}{{\left( t \right)}^{\prime }}}{2\pi }.
\end{aligned}
\end{equation}

In traditional DQD method, the phase noise from the NCO is still existed, but it is counteracted in DMTD method.
This frequency difference ${f}_{r,e}$ includes high-frequency phase noise $\frac{\Delta {{\varphi }_{r}}{{\left( t \right)}^{\prime }}-\Delta {{\varphi }_{e}}{{\left( t \right)}^{\prime }}}{2\pi }$, 
which introduces measurement errors.
However, the frequency difference $m\Delta f$ remains relatively stable over short periods, while the phase noise decreases when averaged.
Thus, by averaging over short periods, the phase noise decreases, and the frequency difference stabilizes:
\begin{equation}
  \label{equation13}
{\overline{{f}_{r,e}}}=m\Delta f+\overline{\frac{\Delta {{\varphi }_{r}}{{\left( t \right)}^{\prime }}-\Delta {{\varphi }_{e}}{{\left( t \right)}^{\prime }}}{2\pi }}.
\end{equation}
 
With averaging, the phase noise diminishes, and the frequency difference approaches $m$, 
providing a stable input for digital feedback algorithms that lock the laser repetition frequency.

Due to inevitable differences between the NCO signal, the measured signal, and the reference signal, 
the phase between them slowly but steadily drifts, causing phase wraps\cite{gdeisat2011one}. 
During frequency calculation, phase unwrapping is required to obtain accurate data. 
Differentiating the phase data can convert it into instantaneous frequency. 
Then averaging the frequency difference can reduce the impact of phase noise.

However, random-access memory (RAM) resources in the FPGA are limited, 
direct averaging over long time periods (e.g., 100 million samples per second at a 100 MHz ADC sampling rate) is impractical. 

\begin{figure}[t]
  \centering
  \includegraphics[width=0.9\linewidth]{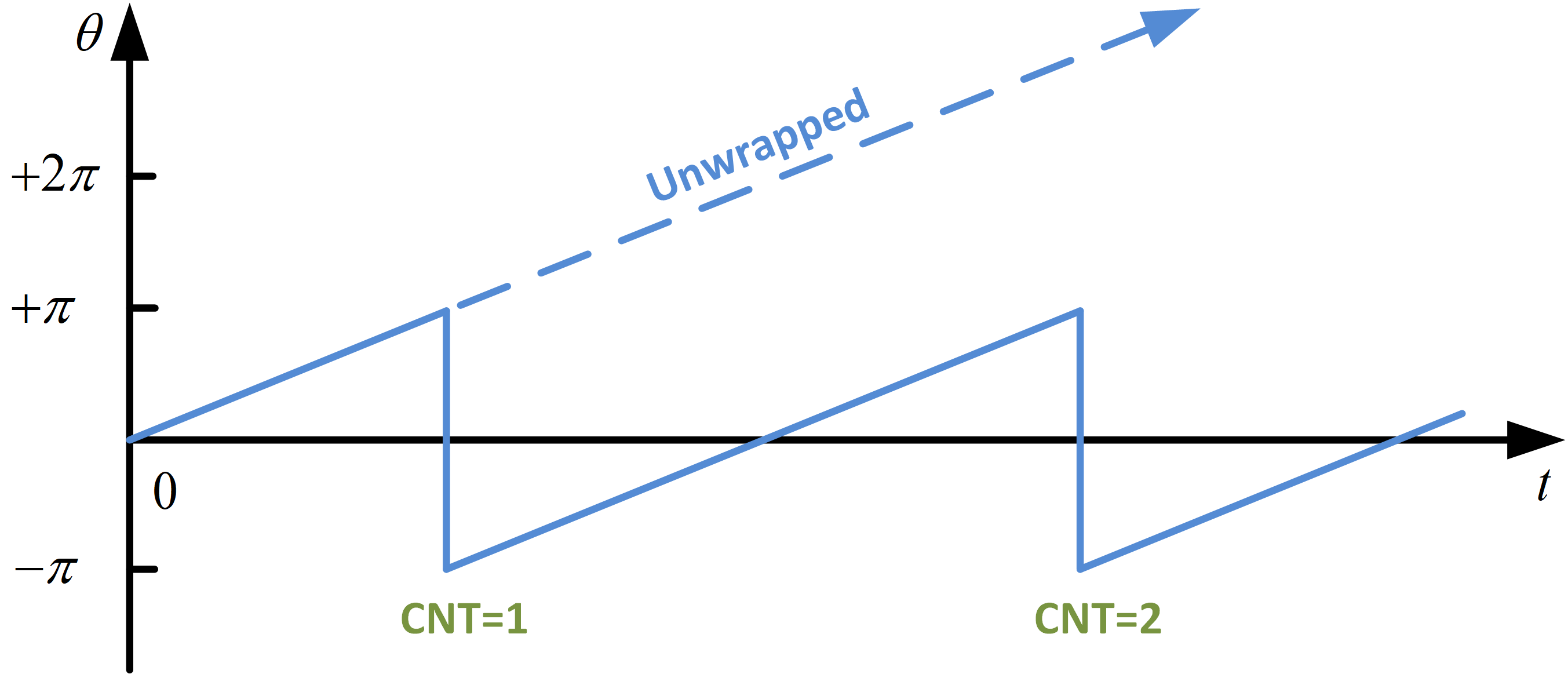}
  \caption{Phase unwrapping diagram. CNT: counter.}
  \label{fig:unwrap}
\end{figure}
Additionally, different feedback devices operate at varying response speeds, 
and it may be necessary to adjust the measurement period to meet specific requirements. 
To address these challenges and minimize phase noise, 
a counter-based phase unwrapping method is employed for frequency calculation and decimation. 
This method requires only a counter (CNT) and minimal RAM, allowing for frequency averaging over a wide range. 
Fig.\ref{fig:unwrap} illustrates the phase unwrapping process, 
where the CNT increments or decrements depending on the direction of the phase wrap. 
By adding the CNT's phase value to the current phase and differentiating over different time intervals, 
frequency measurements can be obtained.

In a digital system, signals consist of discrete sampled points. 
As the sample rate of the ADC is $F_s$, its sample period is $T_s$.
Based on the derived approach above, a series of phase points is obtained within the FPGA, 
denoted as ${\theta(1), \theta(2), \ldots, \theta(n-1), \theta(n)}$. 
The frequency at any given point is expressed as:
\begin{equation}
  \label{equationCNT1}
  f(n) = \frac{\theta(n) - \theta(n-1)}{2\pi T_s}.
\end{equation}

Averaging the frequency over multiple points results in:
\begin{equation}
  \label{equationCNT2}
  \overline{f} = \frac{1}{2\pi (m-1) T_s} \sum_{n=2}^{m} \left( \theta(n) - \theta(n-1) \right)
\end{equation}
where $m-1$ is the length of the measurement period. 
Simplifying this expression yields:
\begin{equation}
  \label{equationCNT3}
  \overline{f} = \frac{\theta(m) - \theta(1)}{2\pi (m-1) T_s}.
\end{equation}

Thus, only the first and last phase values need to be stored in the FPGA to calculate the frequency over the cycle of $m$. 
Using a counter, the unwrapped phase values $\theta_u(1)$ and $\theta_u(m)$ can be determined as follows:
\begin{equation}
  \label{equationCNT4}
  \theta_u(1) = \theta(1)
\end{equation}
\begin{equation}
  \label{equationCNT5}
  \theta_u(m) = \theta(m) + 2\pi \text{CNT}(m).
\end{equation}

Finally, the average frequency difference is expressed as:
\begin{equation}
  \label{equationCNT6}
  \overline{f} = \frac{\theta_u(m) - \theta_u(1)}{2\pi (m-1) T_s} = \frac{\theta(m) - \theta(1) + 2\pi \text{CNT}(m)}{2\pi (m-1) T_s}.
\end{equation}

This counter-based phase unwrapping method allows for frequency measurements over periods ranging from nanoseconds to seconds, 
with adjustable feedback periods to suit different system requirements.

\subsection{Feedback Loop}
In the feedback loop, the amplified frequency difference serves as the error signal for laser stabilization. 
This signal is first processed by an LPF to eliminate high-frequency noise outside the feedback bandwidth. 
The error signal is then passed to a PID controller, 
which computes the difference between the error signal and the settled frequency offset, 
allowing flexible locking of the laser to various frequencies.
The calculated error is converted into an output voltage by a DAC, 
then amplified by a high-voltage amplifier. 
This voltage is applied to a piezoelectric transducer (PZT) to adjust the laser's frequency. 
As all feedback components are digital, the system can be easily configured to support different feedback devices, such as PZTs and servos.

\section{Implementation and Performance} \label{Implementation and Performance}
\subsection{Implementation}
\begin{figure}[t]
  \centering
  \includegraphics[width=0.98\linewidth]{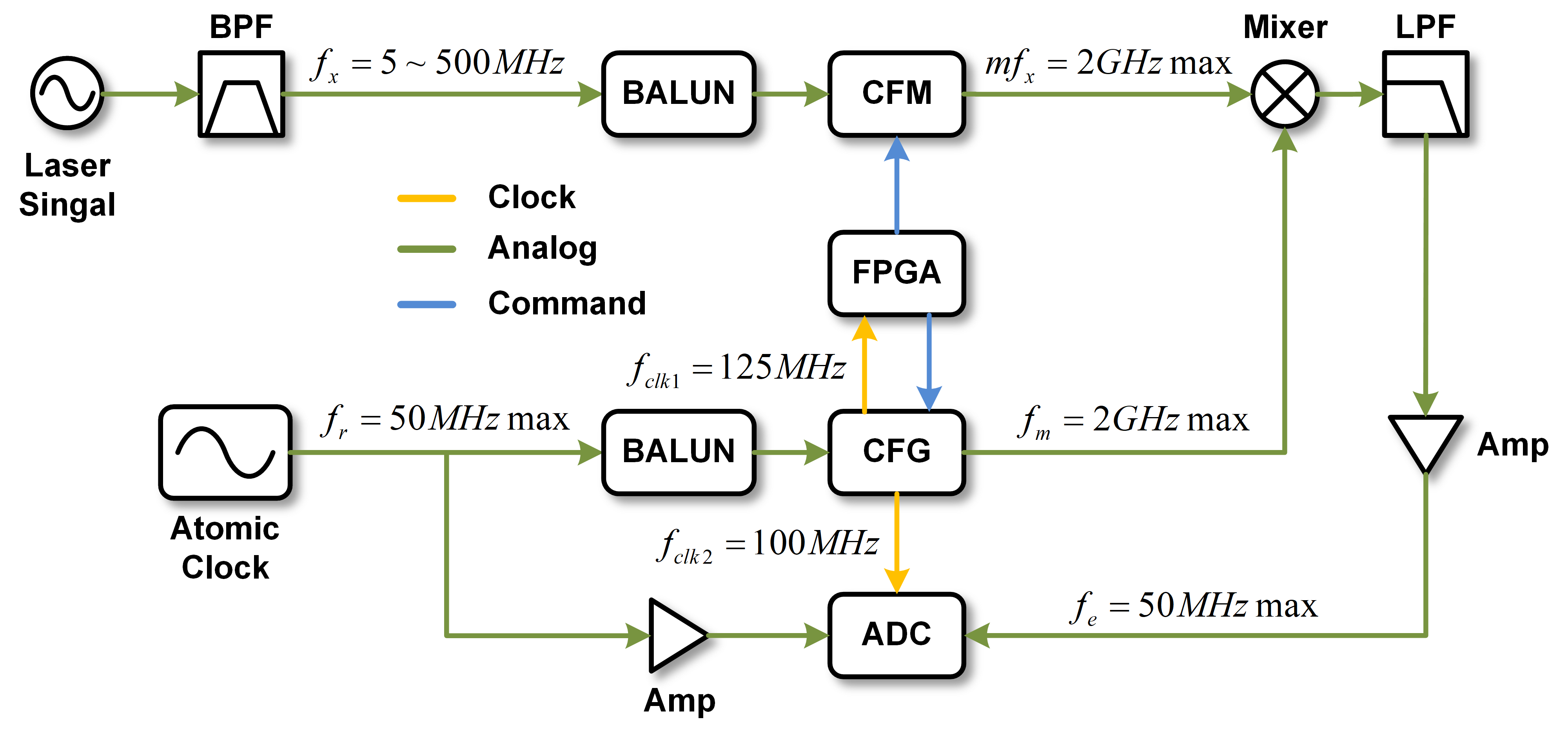}
  \caption{Block diagram of the digitized error amplification module.}
  \label{fig:error_amp}
  \end{figure}
\begin{figure}[t]
  \centering
  \includegraphics[width=0.98\linewidth]{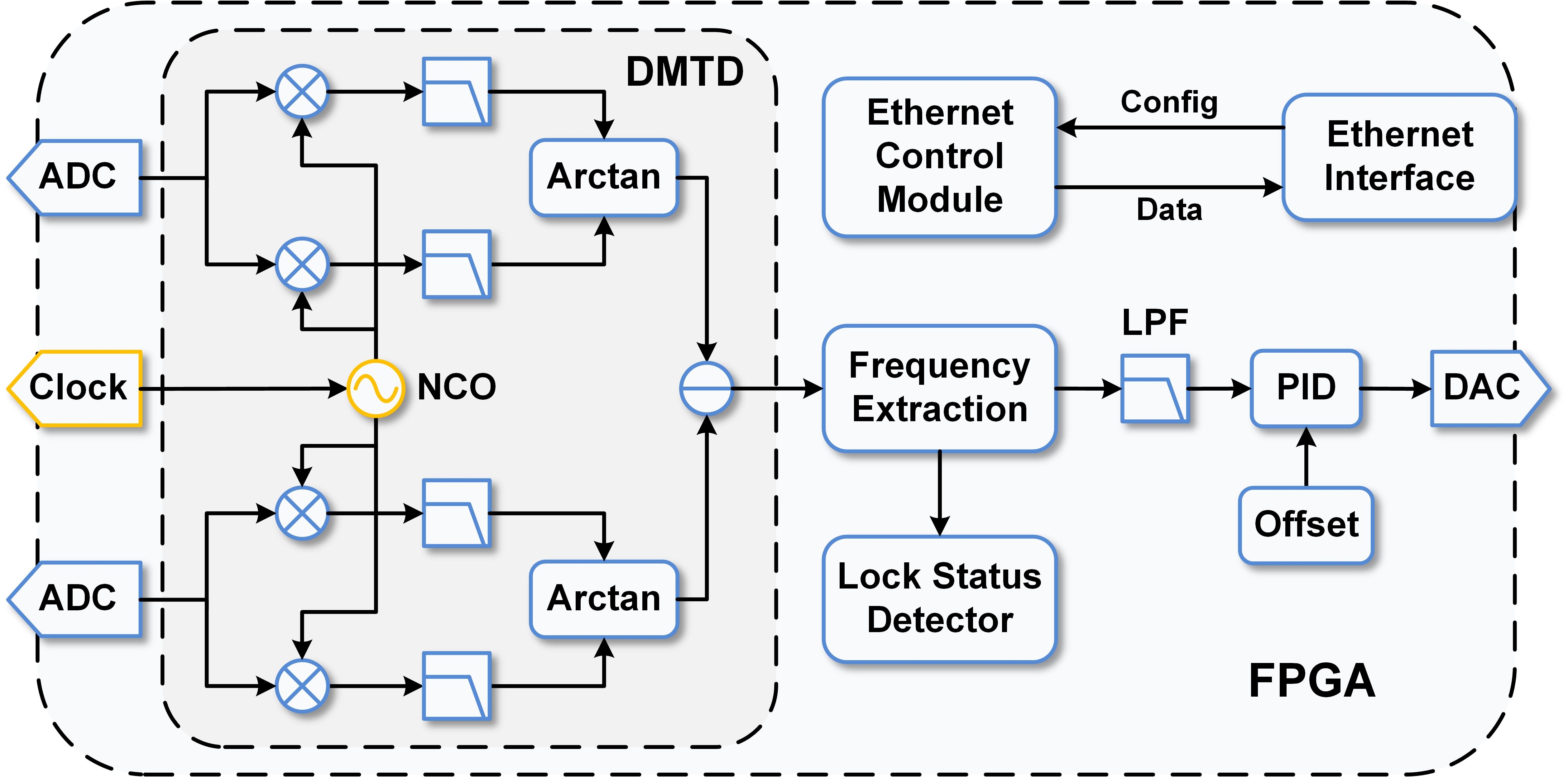}
  \caption{Function modules of FPGA logic firmware.}
  \label{fig:FPGA}
  \end{figure}
  \begin{figure}[t]
    \centering
    \includegraphics[width=0.9\linewidth]{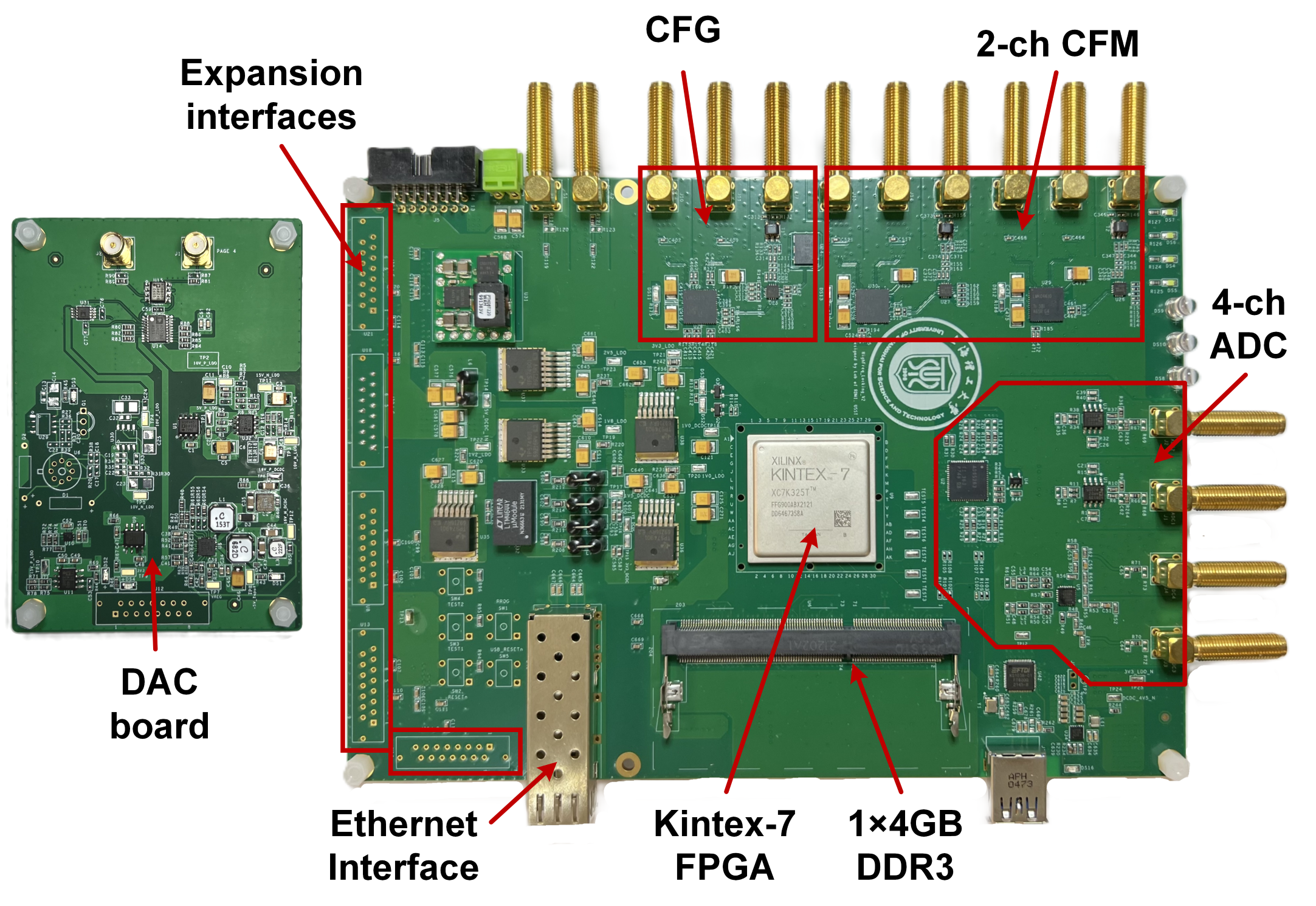}
    \caption{Annotated photograph of the high-speed digital FPGA board.}
    \label{fig:board}
  \end{figure}
The proposed electronic prototype includes an EAM and a high-speed digital FPGA board.
Fig.\ref{fig:error_amp} shows the block diagram of the EAM.
The EAM utilizes the LMK04610 chip from Texas Instruments as the CFG and CFM. 
This chip can handle input signals from 5 to 500 MHz with output frequencies reaching up to 2 GHz, 
which is adequate for various laser repetition frequency. 
BALUNs are employed for signal conversion, minimizing external noise\cite{lewallen1985baluns}. 
The CFG provides clock signals to the FPGA, generates ADC sampling signals, 
and supplies the local oscillator signals for mixing. 
Based on the sampling rate of the system's ADC, the EAM generates signals matching the ADC's sampling frequency.

The function modules of the FPGA logic firmware are shown in Fig.\ref{fig:FPGA}. 
The ADC-based-DMTD, frequency extraction method as well as the feedback loop are deployed in the FPGA for frequency locking.
What's more, a lock status detector module can detect the frequency difference in real-time, 
outputting a locked signal when the frequency error remains within a specified range for a certain period.
The Ethernet interface allows configuration of the EAM and DFLM by computer. 
Configurable parameters include the multiplication factor, CFG output frequency, 
frequency measurement period in the Frequency Extraction module, the frequency offset between the laser and reference, 
low-pass filter parameters, PID parameters, and DAC feedback speed. 
The EAM is adaptable to different laser repetition frequencies, 
and the backend feedback module can be configured to accommodate various feedback devices such as PZTs and servos.

Fig.\ref{fig:board} shows the annotated photograph of the high-speed digital FPGA board.
The digital frequency locking board fits within a $20 \times 15 \times 5$ $cm^3$ volume and can be used to lock two independent lasers.
A Xilinx Kintex-7 FPGA is used as the central control unit, 
with the platform's digital signal control entirely managed by the programmable digital logic within the FPGA. 
The power supply section includes both analog and digital circuits to generate the main power supply. 
Communication between the computer and the FPGA is facilitated through a 1G Ethernet interface. 
The board contains a CFG and two CFMs to support multiple channel frequency stabilization. 
The platform integrates a four-channel high-speed ADC with a resolution of 16 bits and a sampling rate of 100 MS/s. 
A high-precision DAC, connected via an expansion interface, features a resolution of 20 bits and an output frequency of 1 MS/s. 
Multiple external expansion interfaces are available for connecting additional DACs and other devices such as DDS. 
All input and output channels are equipped with 50 $\Omega$ Sub-miniature version A (SMA) terminations. 
A 4-GB DDR3 memory provides sufficient data storage capacity and memory bandwidth.

This system is suitable for locking a single laser and can be expanded to lock dual laser combs by adding more DACs. 
The reprogrammable FPGA chip offers excellent flexibility and scalability to meet various requirements.
By integrating higher-precision ADCs and faster DACs, the system's performance can be further improved. 
Additionally, coordinating the use of high-speed and low-speed DACs may enhance locking performance.

\subsection{DFLM Noise Floor}
\begin{figure}[t]
  \centering
  \includegraphics[width=0.98\linewidth]{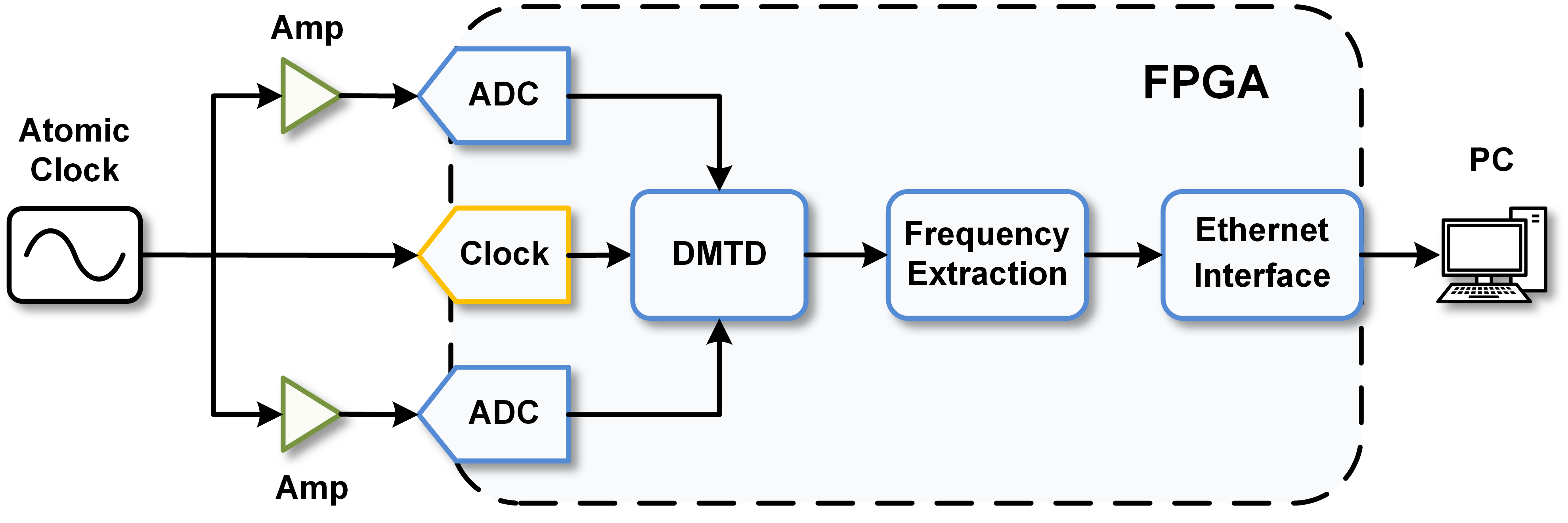}
  \caption{Schematic diagram of DFLM noise floor measurement.}
  \label{fig:noise_floor}
\end{figure}
Fig.\ref{fig:noise_floor} illustrates the setup for measuring the DFLM noise floor. 
A reference source is split using a power divider and fed into both channel of the DFLM module through equal-length RF cables. 
The data is then transmitted via Ethernet Interface to the local PC for stability analysis. 
Since the two input sine wave signals originate from the same source, their phase difference remains at zero\cite{4303427}, 
making this configuration ideal for measuring the system's noise floor. 
The reference source used is a rubidium atomic clock, the SAFRAN LPFRS-01, 
with a nominal frequency of 10 MHz and a stability of $2 \times 10^{-11}$ at 1 second. 
The sampling interval was set to 100 kHz, with a total measurement duration of 400 seconds.

\begin{figure}[t]
  \centering
  \includegraphics[width=0.98\linewidth]{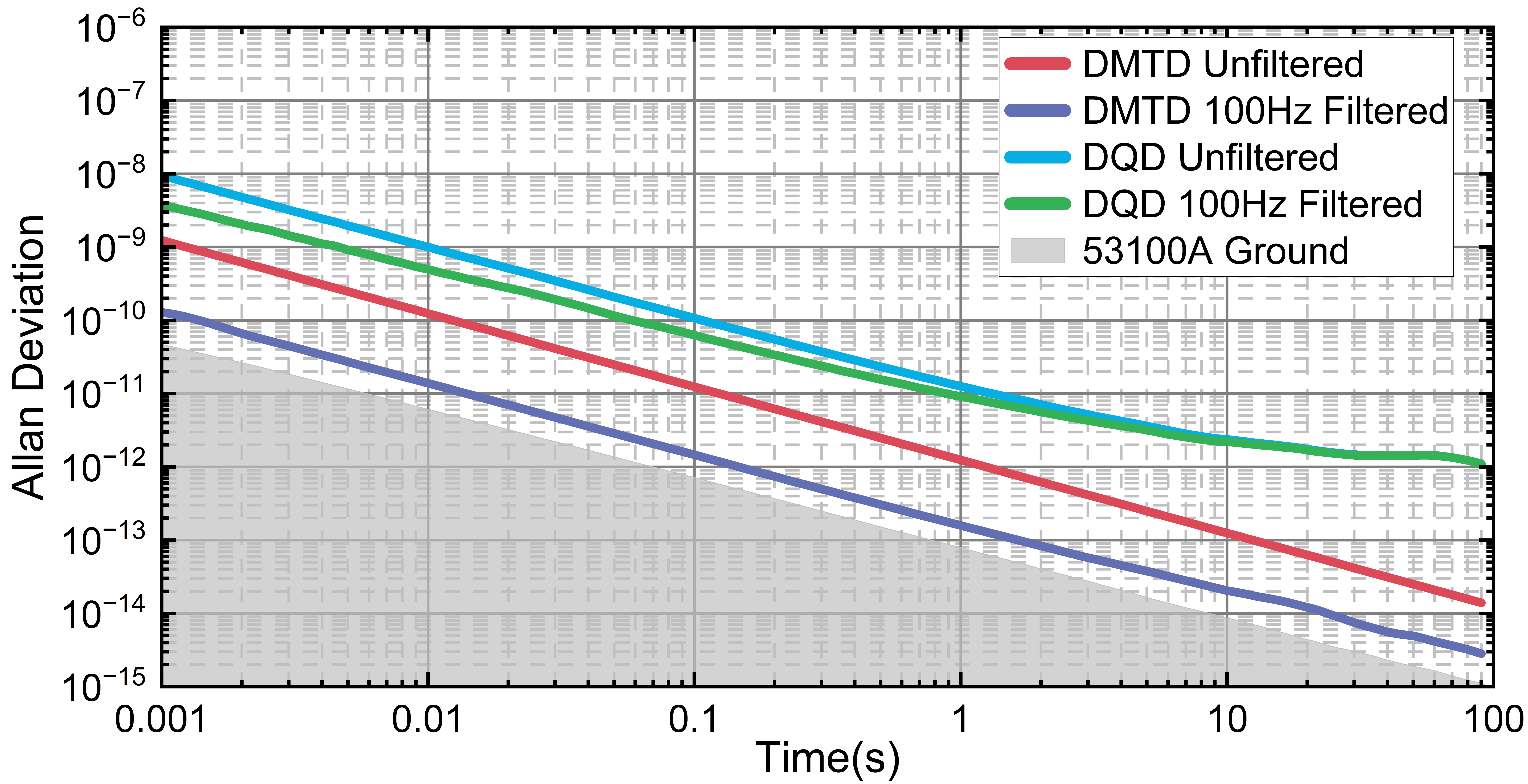}
  \caption{Time deviation in ADC-based-DMTD method and DQD method.}
  \label{fig:noise_floor_compare}
\end{figure}
The precision of the ADC-based-DMTD method described in this paper was compared with the DQD method. 
Both methods use the same external test circuit structure,
but the ADC-based-DMTD frequency extraction module is replaced with the DQD frequency extraction module in DQD noise floor test. 
Fig.\ref{fig:noise_floor_compare} presents the measured DFLM noise floor, 
displaying the results both before and after filtering for both methods. 
The figure shows the unfiltered noise floor and the noise floor after applying a 100 Hz low-pass filter for both methods. 
The ADC-based-DMTD method demonstrated better noise floor both before and after filtering. 
After filtering, the ADC-based-DMTD method achieved a noise floor of approximately $1.58 \times 10^{-13}@1s$,
while the traditional DQD method only reached $9.16 \times 10^{-12}@1s$.
Additionally, the noise floor of the phase noise analyzer (Microchip model 53100A) are also shown in the figure, 
taken at a 1 kHz measurement period with a 500 Hz low-pass filter.
The analyzer's noise floor was $7.96 \times 10^{-14}$ at a gate time of 1s.

The results demonstrate that the DFLM using the ADC-based-DMTD method achieves a noise floor 
comparable to that of a commercial phase noise analyzer, offering more precise frequency measurements.

\subsection{VCO Stability Experiment}
\begin{figure*}[t]
  \centering
  \includegraphics[width=0.95\linewidth]{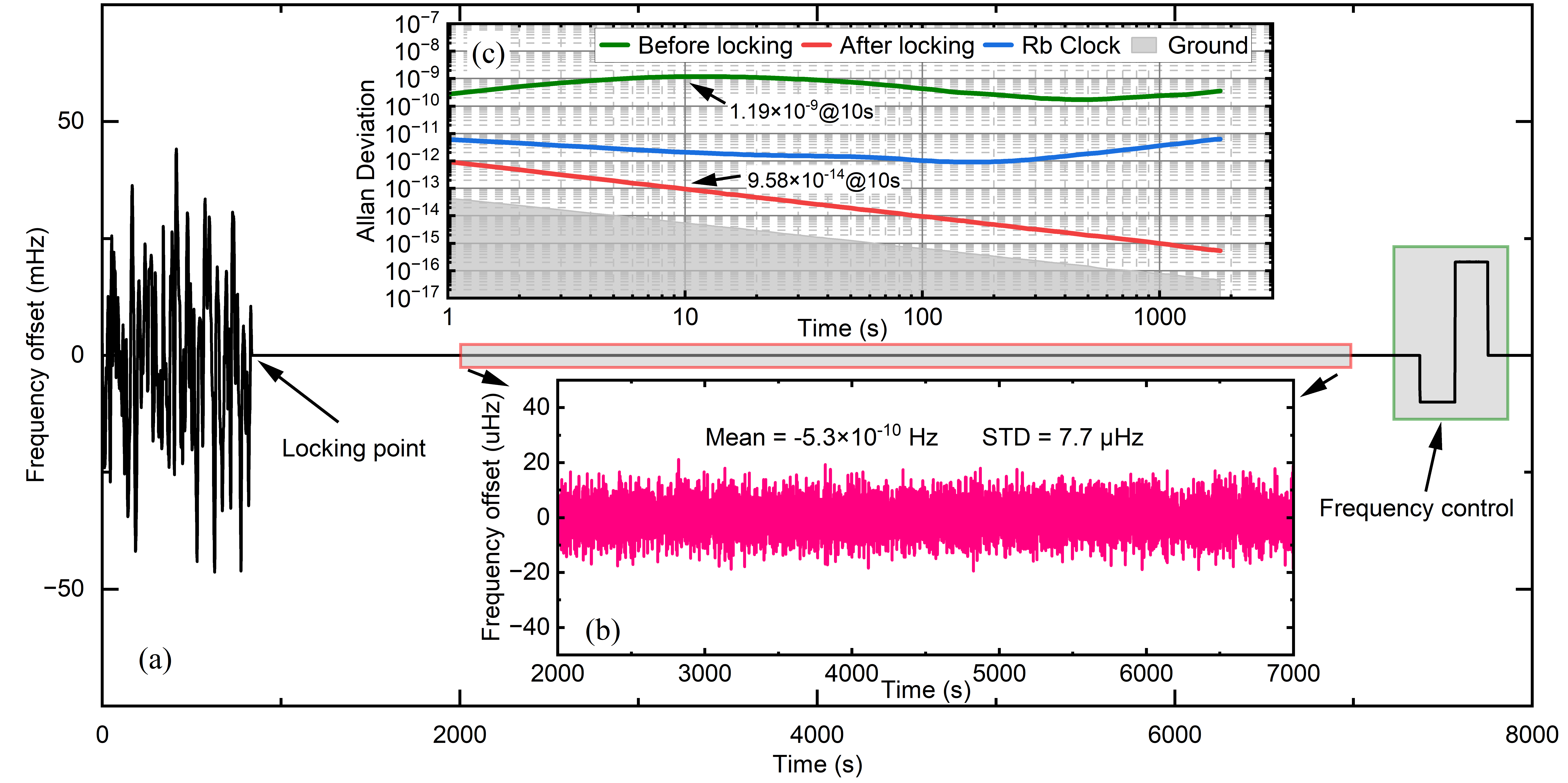}
  \caption{Fig(a) illustrates the frequency offset of the voltage-controlled oscillator (VCO) relative to the reference frequency over 8000 seconds, showing results both before and after locking, as well as the frequency control portion. 
  Fig(b) offers a zoomed-in view of the frequency fluctuations after locking, with a standard deviation (STD) measured at 7.7 \textmu Hz root mean square (RMS). 
  Fig(c) shows the Allan variance before and after locking, with the green line representing the before-locking state and the red line depicting the after-locking state. 
  For comparison, the Allan deviation between two independent rubidium atomic clocks is displayed in blue, 
  while the ground noise of the 53100A phase noise analyzer is shown in gray.}
  \label{fig:VCO}
\end{figure*}
To assess the system's frequency-locking performance, 
we conducted a stability test using a high-stability voltage-controlled oscillator (VCO). 
The VCO effectively simulates laser behavior with its stable output frequency and voltage-controlled adjustability. 
Furthermore, using the VCO minimizes environmental interference, enabling a more controlled evaluation. 
Its superior short-term stability, low long-term stability, 
and high feedback bandwidth make it an optimal choice for demonstrating the system's precision and performance.

The VCO used in this test is the AOCJY1 model from Abracon LLC, with a nominal frequency of 10 MHz, 
a frequency adjustment range of ±100 Hz, and a nominal stability of $1 \times 10^{-9}@1s$. 
In the EAM, a 100x frequency multiplication was applied, amplifying the VCO frequency to 1 GHz. 
After down-conversion, the error signal was obtained with an amplification factor of 100. 
Given the nominal stability of the VCO, 
a 300 Hz low-pass filter was applied to the error signal in the DFLM to eliminate high-frequency noise, 
with a feedback frequency set at 100 kHz.

To evaluate the frequency locking performance and long-term stability, 
a Microchip phase noise analyzer (model 53100A) was used to measure the frequency difference.
Both the reference for frequency locking and the analyzer were connected to a high-stability rubidium atomic clock with a nominal stability of $2 \times 10^{-11}@1s$. 
The gate time of the phase noise analyzer is set to 1 second.
The total measurement time was 6 hours, with 8000 seconds selected for display. 
The frequency offset refers to the offset between the measured frequency and the nominal frequency.

As shown in Fig.\ref{fig:VCO}(a), from 0 to 836 seconds, the VCO operated in free-run mode. 
At 837 seconds, the FPGA initiated frequency locking, successfully locking the VCO to the reference frequency within 1 second, 
maintaining the lock for the remainder of the test. 
At the end of the figure, the system's frequency control capability is demonstrated by switching between two different frequency offsets, 
achieving real-time and precise frequency locking. 
Unlike traditional PLLs, the system allows for arbitrary frequency locking within the allowable precision range, 
offering excellent frequency stability with minimal parameter adjustment required.
Fig.\ref{fig:VCO}(b) provides a zoomed-in view of the frequency fluctuations after locking, with a standard deviation (STD) of 7.7 \textmu Hz RMS. 
Fig.\ref{fig:VCO}(c) presents the Allan deviation before and after locking, with the green line indicating the before-locking state and the red line showing the after-locking state. 
For comparison, the Allan deviation between two independent rubidium atomic clocks is displayed in blue, 
while the ground noise of the 53100A phase noise analyzer is shown in gray.
With locking enabled, the VCO achieved a short-term Allan deviation at 1 second of $9.48 \times 10^{-13}$, 
a significant improvement over the $2.73 \times 10^{-10}$ observed before locking. 
When the averaging time comes to 10 seconds, the VCO exhibited a stability of $1.19 \times 10^{-9}$ before locking, 
which improved to $9.58 \times 10^{-14}$ after locking. 
Even highly stable atomic clocks exhibit inherent random frequency fluctuations. 
With the VCO locked closely to the reference atomic clock, 
its frequency precisely followed the reference's fluctuations, 
resulting in a measurement stability that surpassed the stability of two independent atomic clocks 
and approached the analyzer's baseline noise level.

The locked VCO demonstrated exceptional stability, confirming the system's accuracy and reliability.

\subsection{Laser Stability Experiment}
\begin{figure*}[!t]
  \centering
  \includegraphics[width=0.95\linewidth]{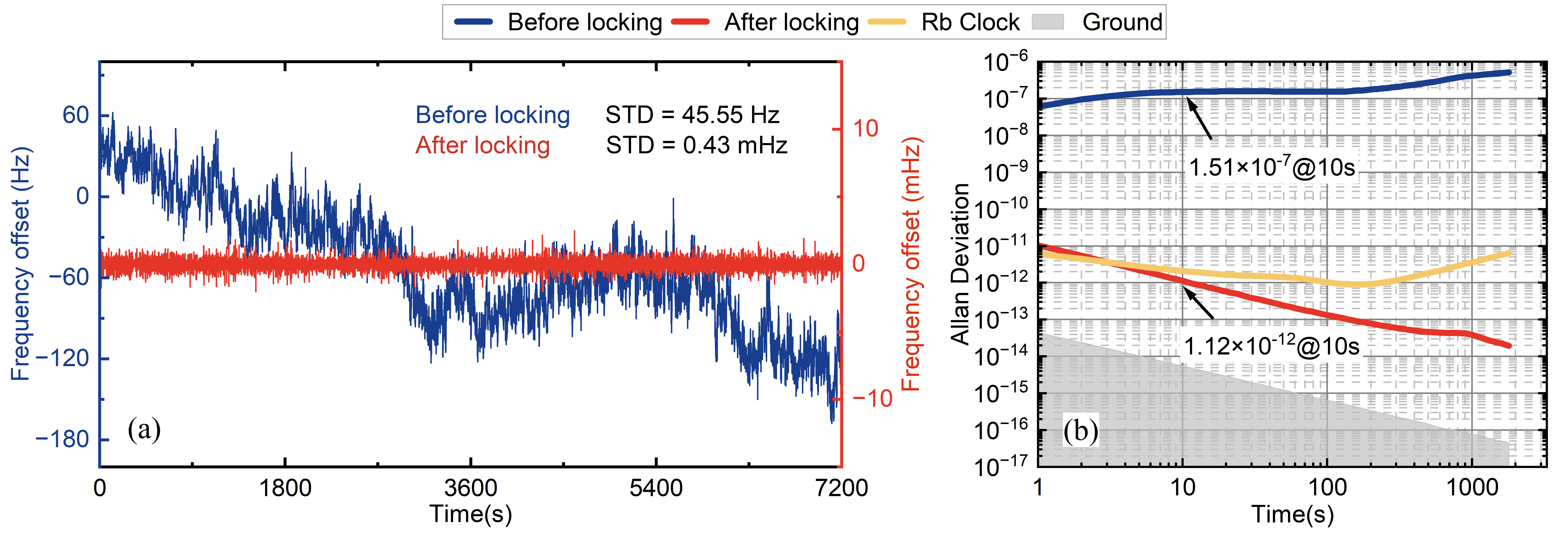}
  \caption{(a) Laser repetition frequency offset over 2 hour period. (b) Allan deviation before and after locking.}
  \label{fig:laser_lock}
\end{figure*}
\begin{figure}[!t]
  \centering
  \includegraphics[width=0.9\linewidth]{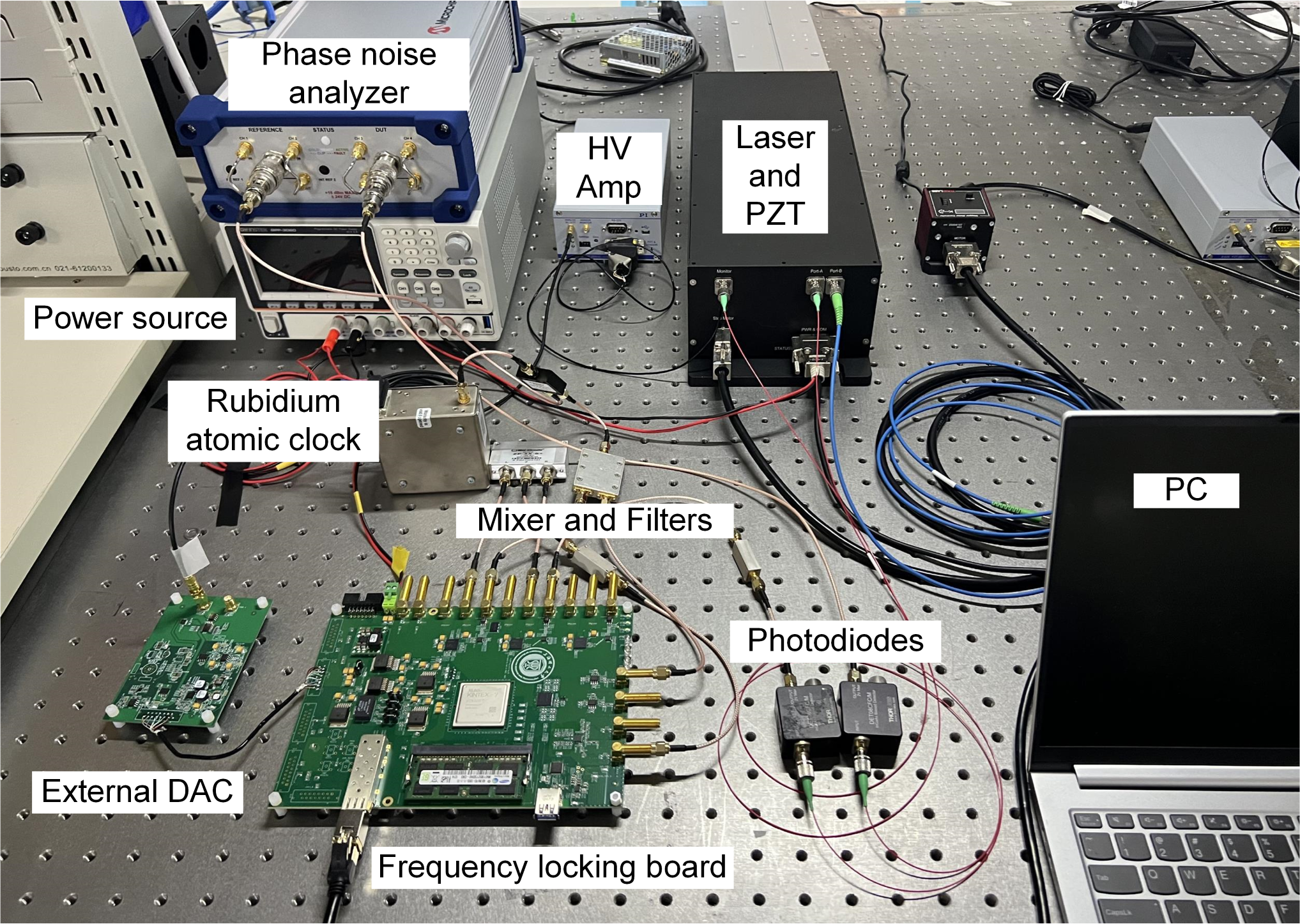}
  \caption{Scene of the laser frequency locking experiment.}
  \label{fig:laser}
\end{figure}
In a standard laboratory environment, 
a laser frequency locking experiment was performed using a custom-built mode-locked femtosecond fiber laser.
Fig.\ref{fig:laser} shows the scene of the laser frequency locking experiment.
The laser uses a semiconductor saturable absorber mirror (SESAM) and a 9-shaped resonator 
to generate a femtosecond pulse seed source, achieving passive mode-locking. 
The repetition rate is tuned by adjusting the cavity length using a PZT.
The laser operates at a center wavelength of 1550 nm, with a repetition rate of 50 MHz and a tunable range of ±400 Hz, 
delivering an average optical power of 100 mW. 
Two 4 mW optical signals were extracted via a fiber coupler and connected to PDs for optical-to-electrical conversion. 
One signal was directed to the system for frequency locking, 
while the other was sent to a commercial phase noise analyzer (model 53100A) to measure the frequency difference.

As the laser used in the experiment has a repetition frequency of 50 MHz, a 20x frequency multiplication was applied in the EAM,
which amplifies the frequency error by a factor of 20.
In the DFLM, a 500 Hz LPF was applied to the error signal to remove high-frequency noise, 
and the filtered signal was then sent through a DAC to control the laser repetition frequency by adjusting the PZT. 
The feedback frequency was set to 1 kHz, which was limited by the PZT.

The system was locked using the rubidium atomic clock mentioned earlier as the high stability reference source, 
which has a nominal stability of $2 \times 10^{-11}@1s$. 
The same phase noise analyzer (model 53100A) was used to measure the frequency difference 
and the rubidium atomic clock was also used as the reference source for the phase noise analyzer.
The gate time was initially set to 0.001 seconds due to the large free-running frequency difference, 
and later adjusted to 1 second after locking.

Data were collected for two hours both before and after locking. The results are displayed in Fig.\ref{fig:laser_lock}. 
The Fig.\ref{fig:laser_lock}(a) shows the laser's frequency offset before and after locking.
The STD of the laser before locking is 45.55 Hz RMS, while the frequency fluctuation stabled at 0.43 mHz RMS after locking,
demonstrating a significant reduction in frequency fluctuations once locked. 
On the right, the Fig.\ref{fig:laser_lock}(b) presents the Allan deviation, 
where the laser's short-term Allan deviation at 1 second improved from $6.05 \times 10^{-8}$ before locking to $1.04 \times 10^{-11}$ after. 
For the averaging time of 10 seconds, the laser achieved a stability of $1.12 \times 10^{-12}$ after locking, 
a substantial improvement from the before-locking stability of $1.51 \times 10^{-7}$.
Additionally, the Fig.\ref{fig:laser_lock}(b) shows the noise floor of the phase noise analyzer used in the measurement, 
as well as results obtained by measuring two different atomic clocks with the same nominal stability. 

\section{Conclusion} \label{Conclusion}
In this paper, we have presented a high-precession, compact electronic prototype for laser repetition frequency locking. 
The system utilizes a PLL-based CFG and CFM in the EAM to amplify frequency errors. 
Combined with a high-speed ADC-based DMTD method, 
it enables precise frequency measurement and reliable frequency locking with minimal parameter adjustments.
The FPGA-based architecture as well as the EAM enables flexibility and adaptability, 
allowing the system to be reconfigured for various feedback devices and laser frequencies.

The results from the VCO Stability experiment confirms the system's high precision frequency locking capability, 
with an Allan deviation of $9.58 \times 10^{-14}$ at 10 second and an STD of 7.7 \textmu Hz RMS after locking,
making it highly suitable for applications requiring long-term stability. 
The Laser Stability experiment demonstrates the system's capability to significantly reduce frequency fluctuations, 
achieving an Allan deviation improvement from $1.51 \times 10^{-7}$ to $1.12 \times 10^{-12}$ at 10 second and an STD of 0.43 mHz RMS after locking. 

By incorporating advanced components such as higher-speed, higher-precision ADCs and DACs, 
or lower-noise PLLs, the system can achieve even greater measurement precision and enhanced feedback bandwidth.

All these demonstrate that the system for laser repetition frequency locking offers a precise, 
reconfigurable, and compact solution for laser frequency stabilization, 
with significant potential for future enhancements in both hardware and software.

\section*{Acknowledgments}
The authors would like to thank Qing Sun at National Institute of Metrology for useful discussions.

 \bibliographystyle{IEEEtran}
\bibliography{IEEEabrv,reference}


 




\vfill

\end{document}